\documentclass[11pt]{article}
\usepackage{osid,graphicx}                         %
\title{On the reconstruction of diagonal elements of density matrix of quantum optical
states by on/off detectors. }
\author{Giorgio Brida, Marco Genovese, Marco Gramegna 
\\{\footnotesize\it I.N.RI.M., Strada delle Cacce 91,
10135 Torino, Italy \& genovese@inrim.it}\\[2ex]
Matteo G. A. Paris \\{\footnotesize\it Dipartimento di
Fisica dell'Universit\`a di Milano, Italia \&
matteo.paris@unimi.it} \\[2ex]
E. Predazzi \\{\footnotesize\it Dip. Fisica Teorica, Univ. Torino and INFN, 
via P. Giuria 1, 10125 Torino, Italy }\\[2ex]
E. Cagliero \\{\footnotesize\it CNISM, sezione Univ. 
Torino, C.so Raffaello 1, 10125 Torino,
Italy }}
\begin{document}
\maketitle
\begin{abstract}
We discuss a scheme for reconstructing experimentally the diagonal
elements of the density matrix of quantum optical states.
Applications to PDC heralded photons, multi-thermal and attenuated
coherent states are illustrated and discussed in some details.
\end{abstract}
\section{Introduction}
The reconstruction of diagonal elements of density matrix of
quantum optical states, i.e. of their photon statistics, is
fundamental for various applications ranging from quantum
information \cite{zei} to the foundations of quantum mechanics
\cite{MG} and quantum optics \cite{man}. Nevertheless, at the
moment, photo-detectors well suited for this purpose are not
available. Indeed, the choice of a detector with internal gain
suitable for the measurement is not trivial when the flux of the
photons to be counted is such that more than one photon is
detected in the time-window of the measurement, which is set by
the detector pulse-response, or by an electronic gate on the
detector output, or by the duration of the light pulse. One needs
a congruous linearity in the internal current amplification
process: each one of the single electrons produced by the
different photons in the primary step of the detection process
(either ionization or promotion to a conduction band) must
experience the same average gain and this gain must have
sufficiently low spread. The fulfillment of both requisites is
necessary for the charge integral of the output current pulse to
be proportional to the number of detected photons. Few example
exist of photo-detectors that can operate as photon counters and
each one has some drawback. Among these, PhotoMultiplier Tubes
(PMT's) \cite{burle} and hybrid photodetectors \cite{NIST} present
a low quantum efficiency, since the detection starts with the
emission of an electron from the photocathode. Solid state
detectors with internal gain, in which the nature of the primary
detection process ensures higher efficiency, are still under
development. Highly efficient thermal photon counters have also
been used, though their operating conditions are still extreme
(cryogenic conditions) to be suited for common use \cite{xxx,
serg}.
\par
The advent of quantum tomography did provide an alternative method
to measure photon number distributions \cite{mun}. However, this
scheme, which was experimentally applied to several quantum states
\cite{raymerLNP}, needs the implementation of homodyne detection,
which in turn requires the appropriate mode matching of the signal
with a suitable local oscillator at a beam splitter. Such mode
matching, generally a not extremely simple scheme to be realized,
is a particularly challenging task in the case of pulsed optical
fields.
\par
On the other hand, the photodetectors  usually employed in quantum
optics, such as Avalanche PhotoDiodes (APD's) operating in the
Geiger mode \cite{rev, serg}, seem to be by definition useless as
photon counters. They are the solid state devices, which present
the highest quantum efficiency and the greatest stability of
internal gain. However, they have the obvious drawback that the
breakdown current is independent of the number of detected
photons, which in turn cannot be determined. The outcome of these
APD's is either "off" (no photons detected) or "on" {\em i.e.} a
"click", indicating the detection of one or more photons.
Actually, such an outcome can be provided by any photodetector
(PMT, hybrid photodetector, cryogenic thermal detector) for which
the charge contained in dark pulses is definitely below the output
current pulses corresponding to the detection of at least one
photon. Note that for most high-gain PMT's the anodic pulses
corresponding to no photon detected can be easily discriminated by
a threshold from those corresponding to the detection of one or
more photons.
\par
Nevertheless, recently the possibility of reconstructing photon
distribution through on/off detection at different efficiencies
has been analyzed \cite{mogy} and its statistical reliability
investigated in some details \cite{pcount}. In addition, the case
of few and small values of $\eta$ \cite{ar} has been addressed. A
first experimental application of the method of \cite{ar,pcount}
has been presented in \cite{nos} showing the very interesting
potentialities of this scheme. In short the method used in ref.
\cite{nos} is based on the measurement of on/off detection
frequencies for a certain optical field when varying the quantum
efficiency of the system, i.e. in practice by interposing
calibrated neutral filters on the optical path. The beauty of this
scheme with respect to alternative ones resides indeed in the
extreme simplicity  that can allow an extensive application of it
to test optical fields in various applications.
\par
With the purpose to refine these initial studies, in this paper we
present a detailed analysis of the possibility of reconstructing
experimentally diagonal elements of density matrix of quantum
optical states and of precisely estimating their uncertainties.
Various examples will be considered, as PDC heralded photons,
multi-thermal and attenuated coherent states.
\section{Theoretical methods}
The statistics of the "no-click" and "click" events from an on/off
detector, assuming no dark counts, is given by
\begin{eqnarray}
p_0(\eta) &=& \sum_n (1-\eta)^n \varrho_n\: \label{p0}\:,
\end{eqnarray}
where $\varrho_n =\langle n | \varrho | n \rangle$ is the photon
distribution of the quantum state with density matrix $\varrho$
and $\eta$ is the quantum efficiency of the detector, {\em i.e.}
the probability of a single photon to be revealed.  At first sight
the statistics of an on/off detector appears to provide quite a
scarce piece of information about the state under investigation.
However, if the statistics about $p_0(\eta)$ is collected for a
suitably large set of efficiency values then the information is
sufficient to reconstruct the whole photon distribution
$\varrho_n$ of the signal, upon a suitable truncation of the
Hilbert space.
\par
The procedure consists in measuring a given signal by on/off
detection using different values $\eta_\nu$ ($\nu=1,...,K$) of the
quantum efficiency.  The information provided by experimental data
is contained in the collection of frequencies $f_{\nu} =
f_0(\eta_\nu) = n_{0\nu}/n_\nu$ where $n_{0\nu}$ is the number of
"no click" events and $n_\nu$ the total number of runs with
quantum efficiency $\eta_\nu$.  Then we consider expression
(\ref{p0}) as a statistical model for the parameters ${\varrho_n}$
to be solved by maximum-likelihood (ML) estimation. Upon defining
$p_\nu\equiv p_0(\eta_\nu)$ and $A_{\nu n} = (1-\eta_{\nu})^n$ we
rewrite expression (\ref{p0}) as $p_{\nu} = \sum_{n} A_{\nu n}
\varrho_n$. Since the model is linear and the parameters to be
estimates are positive (LINPOS problem), the solution can be
obtained by using the Expectation-Maximization algorithm (EM)
\cite{EMalg}. By imposing the restriction $\sum_n \varrho_n = 1$,
we obtain the iterative solution
\begin{equation}
\varrho_n^{(i+1)} = \varrho_n^{(i)}\sum_{\nu=1}^K \frac{A_{\nu
n}}{\sum_m A_{\nu m}}
\frac{f_{\nu}}{p_{\nu}[\{\varrho_n^{(i)}\}]}\: \label{ems}\:
\end{equation}
where $p_{\nu}[\{\varrho_n^{(i)}\}]$ are the probabilities
$p_{\nu}$, as calculated by using the reconstructed distribution
$\{\varrho_n^{(i)}\}$ at the $i$-th iteration. As a measure of
convergence we use the total absolute error at the $i$-th
iteration $\varepsilon^{(i)} =\sum_{\nu=0}^K \left| f_\nu- p_\nu
[\{ \varrho_n^{(i)}\}]\right|$ and stop the algorithm as soon as
$\varepsilon^{(i)}$ goes below a given level. The total error
measures the distance of the probabilities $p_\nu [\{
\varrho_n^{(i)}\}]$, as calculated at the $i$-th iteration, from
the actual experimental frequencies. As a measure of accuracy we
adopt the fidelity $G^{(i)} = \sum_n \sqrt{\varrho_n \:
\varrho_n^{(i)}}$ between the reconstructed distribution and the
theoretical one.
\section{Confidence intervals}
Our reconstruction of the photon statistics is based on the ML
algorithm, which by an iterative method reaches the $\varrho_n$
best reproducing the experimental data, {\em i.e} the statistics
of 'no-clicks' $f_\nu$. The confidence interval on the ML
determinations may be thus evaluated according to the following
argument. Due to the fact that ML methods are unbiased at
convergence  we have $p_\nu \{ \varrho_n^\infty \} \equiv f_\nu$.
If in practice we stop  at a given iteration $L$ we have
fluctuations in the reconstructed $\{\varrho_n\}$ and, in turn,
$p_\nu \{ \varrho_n^L \} \neq f_\nu$. The fluctuations in the
single components $\varrho_n$ leads to
\begin{equation} p_\nu \{ \varrho_n^L \} = f_\nu + \frac{\partial
p_\nu}{\partial \varrho_n} \: (\varrho_n - \varrho_n^*)
\end{equation} where we have denoted by $\varrho_n^*$ the {\em
true} value of the distribution. As a consequence we may write
$\delta\varrho_n^\nu = |p_\nu\{\varrho_n^L\}-f_\nu |/A_{\nu n}$
and, averaging over the different values of the quantum
efficiency, \begin{equation} \delta\varrho_n \simeq \frac{1}{K}
\sum_{\nu=1}^K \frac{|p_\nu\{\varrho_n^L\}- f_\nu |}{A_{\nu n}}
\end{equation}
\begin{figure}[h]
\begin{tabular}{ccc}
\includegraphics[width=0.3\textwidth]{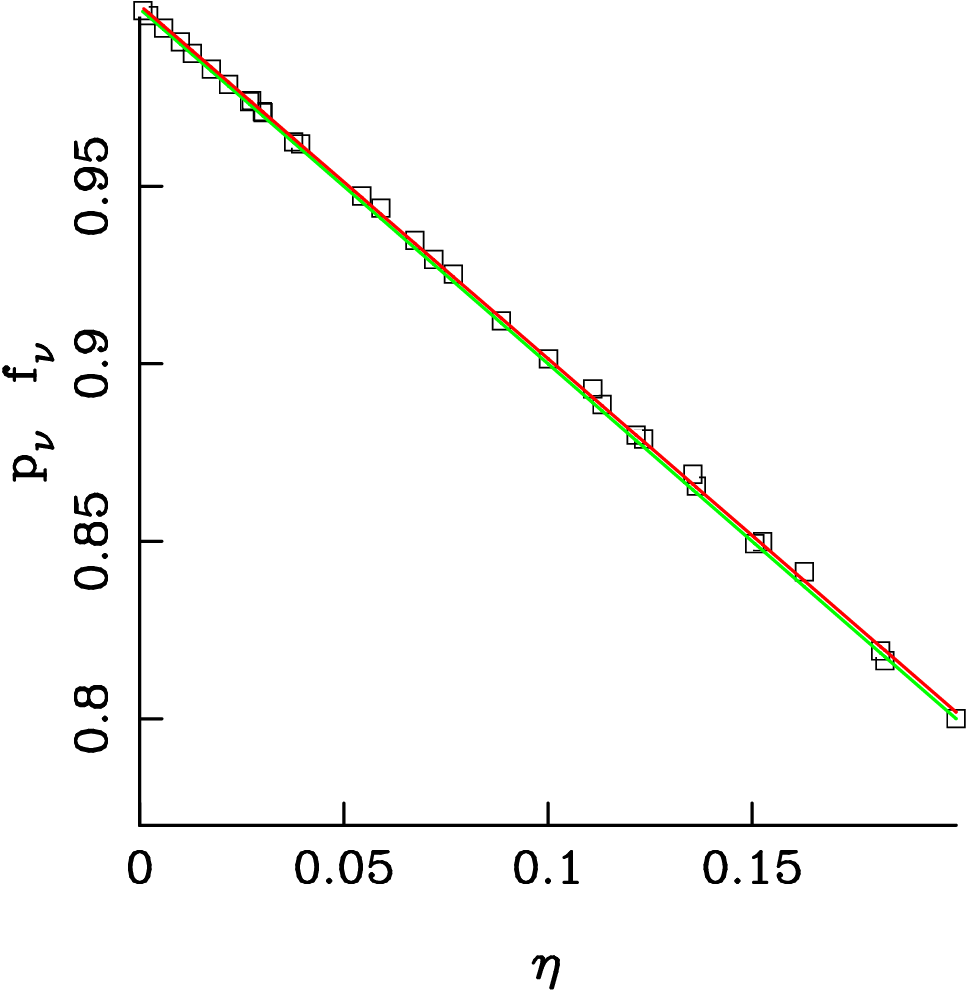}
&
\includegraphics[width=0.3\textwidth]{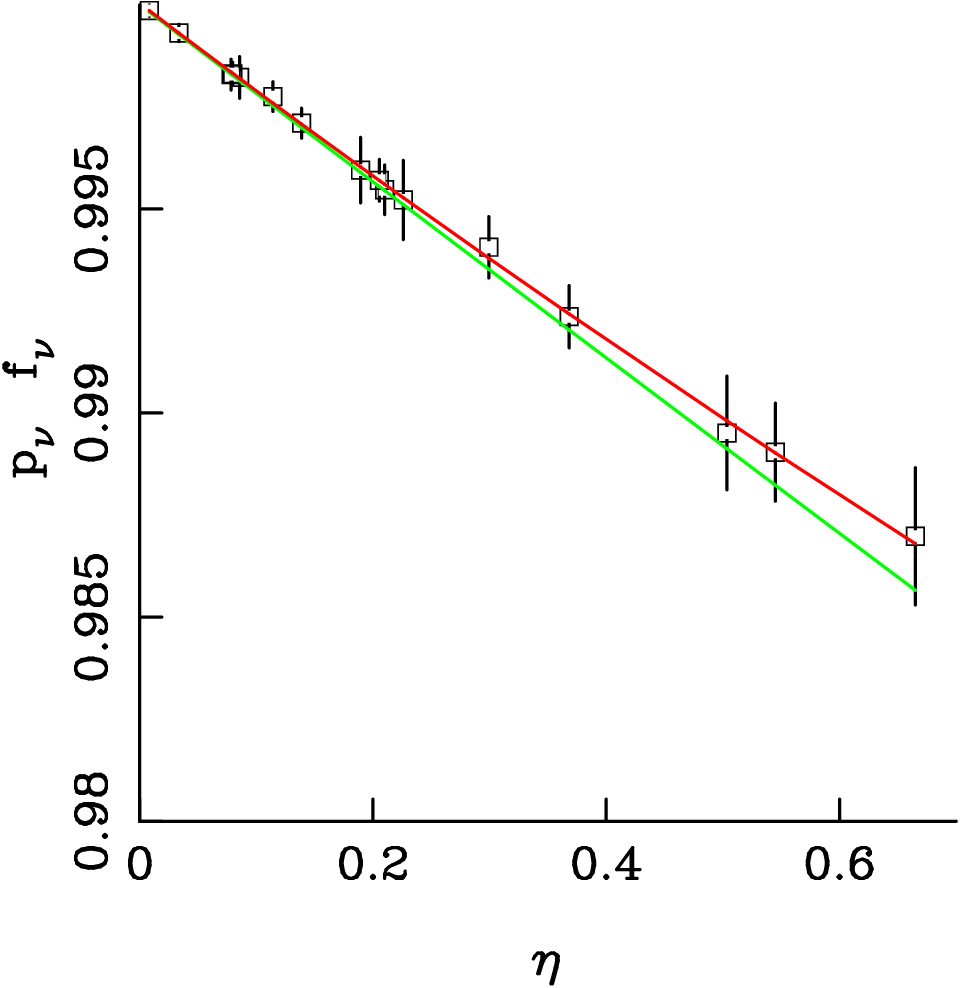}
\\
\includegraphics[width=0.3\textwidth]{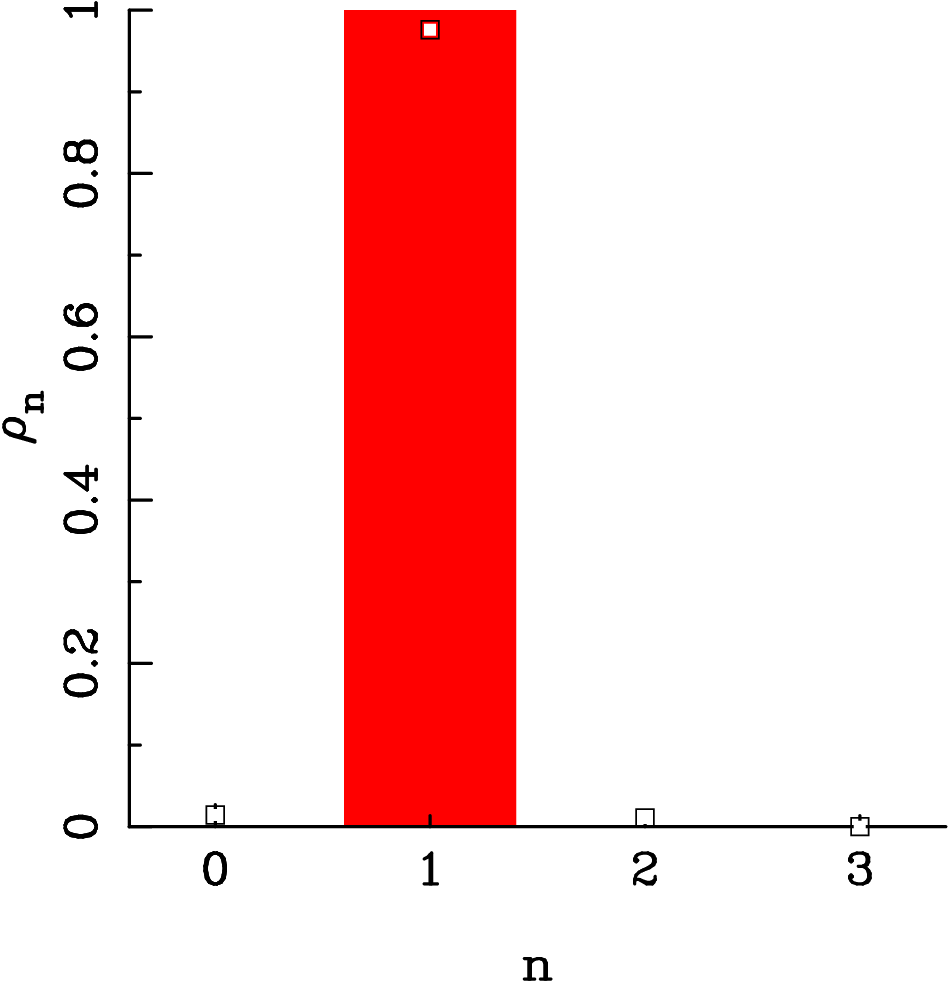}
&
\includegraphics[width=0.3\textwidth]{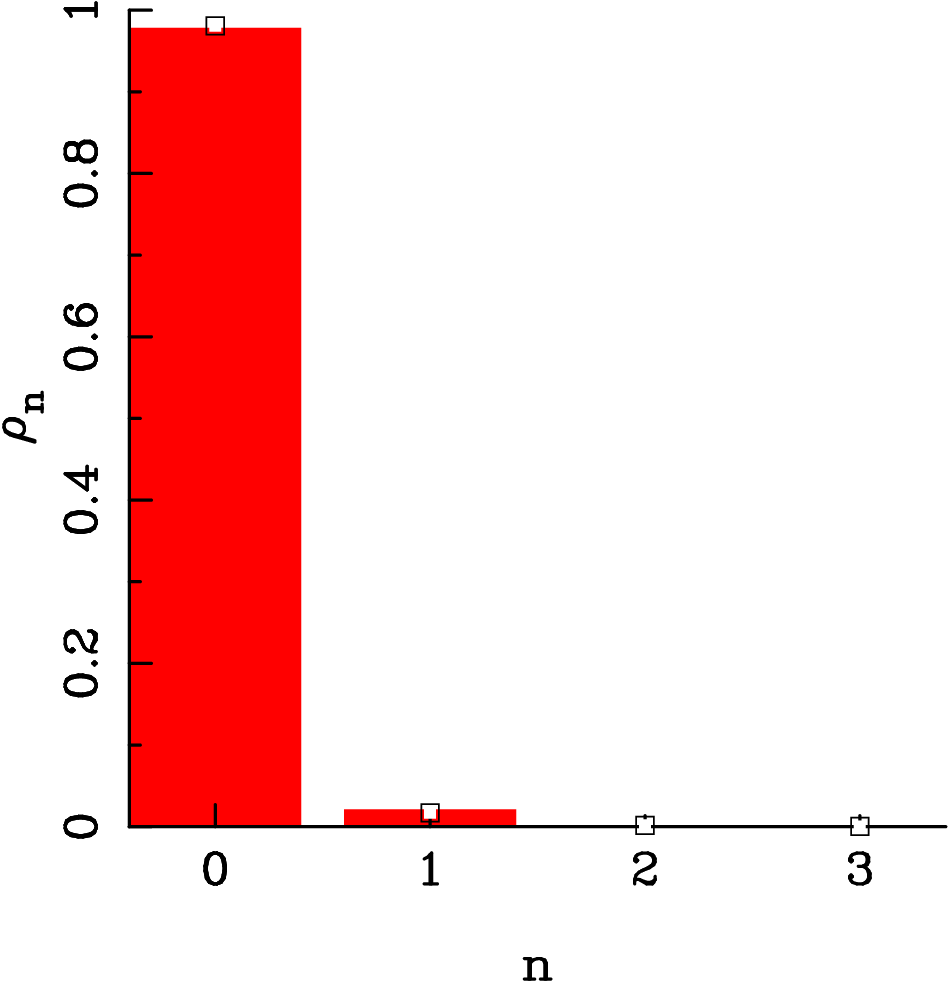}
\end{tabular}
\caption{Reconstruction of statistics of quantum optical states.
In the highest part: Experimental frequencies $f_{\nu}$ of
no-click events compared with reconstructed ones (red  curve) and
theoretical model for the state (black  curve). In the lowest
part, reconstructed probabilities compared with data. a) PDC
heralded photon (produced in type II PDC cw-regime). b) Strongly
attenuated coherent state (He-Ne laser beam)\label{f:fig}}
\end{figure}
\section{Experimental data}
In the following we describe the reconstruction of photon
statistics of several optical states. We begin by considering 
Parametric Down Conversion (PDC) heralded photons.
In our set-up, a pair of correlated photons is generated by
pumping a  $\beta$-Barium-Borate (BBO) crystal with a continuous
wave (cw) Argon ion laser beam (351 nm) in collinear geometry. The
crystal has been cut for producing type II PDC, i.e. photons with
orthogonal polarisation. After having separated the photons of the
pair by means of a polarizing beam splitter, the detection of one
of the two by a silicon avalanche photodiode detector
(SPCM-AQR-15, Perkin Elmer) was used as an indication of the
presence of the second photon in the other channel, namely a
window of 4.9 ns was opened for detection in arm 2 in
correspondence to the detection of a photon in arm 1. The arm 2
"heralded photon" was then measured by a silicon avalanche
photodiode detector (SPCM-AQR-15, Perkin Elmer) preceded by an
iris and an interference filter (IF) at 702 nm, 4 nm Full Width
Half Maximum (FWHM), inserted with the purpose of reducing the
stray light. More in detail, the output of the first detector
started a ramp on a Time to Amplitude Converter (TAC) that
received the second detector signal as stop. The TAC output was
then addressed to a multichannel analyser (MCA) and a window of
4.9 ns was set by selecting only the respective channels of MCA.
The quantum efficiency of the arm 2 detection apparatus (including
IF and iris) was measured to be $20  \%$ by using the PDC
calibration scheme (see \cite{pdccs}). Lower quantum efficiencies
were simulated by inserting calibrated neutral optical filters on
the optical path.

A comparison of the observed frequencies $f_{\nu}$ with the
theoretical curve ($1- \eta_{\nu}$) is presented in the inset of
Fig. 1. In the figure evaluated uncertainties are also shown.

The photon distribution has been reconstructed using $K=34$
different values of the quantum efficiency from $\eta_{\nu}\simeq
0$ to $\eta_\nu\simeq 20\%$ with $n_\nu=10^7$ runs for each
$\eta_\nu$. Results at iteration $i=10^7$ are shown in the figure.
As expected the PDC heralded  photon state largely agrees with a
single photon Fock state. The estimated relative uncertainty for
the $\rho_{1}$ element is 0.1 \%, confirming the precision of the
method. One can notice that also a small two photon component and
a vacuum one are observed. The $\rho_2$ contribution is expected,
by estimating the probability that a second photon randomly enters
the detection window, to be $1.85 \%$ of $\rho_1$, in agreement
within uncertainty  with what observed. A non zero $\rho_0$ is
also expected due to background. This quantity can be evaluated to
correspond to $(2.7 \pm 0.2) \%$ by measuring the counts when the
polarization of the pump beam is rotated to avoid generation of
parametric fluorescence. Also this estimate is in good agreement
with the reconstructed $\rho_0$.

As an example of application of the reconstructed diagonal
elements of the density matrix, a clear test of non-classical
nature of this optical field can then be performed by using the
evaluated probabilities of finding n photons, $p_n$, to estimate
the Klyshko \cite{kl} parameter $ K_n = (n+1) {p_{n-1} p_{n+1}
\over n p_n^2} $, which is always larger than unity for classical
fields. In our case we obtain $ K_1 = (3.2 \pm 0.4) \cdot
10^{-4}$.

 A further reconstruction of diagonal
elements of density matrix has concerned a strongly attenuated
coherent state. This state has been produced by a He-Ne laser beam
attenuated to photon-counting regime  by insertion of neutral
filters.  Also in this case the same silicon avalanche photodiode
detector was used. The distribution has been reconstructed with
$K=15$ different values of the quantum efficiency from
$\eta_{\nu}\simeq 0$ to $\eta_\nu\simeq 66\%$ with $n_\nu=10^7$
runs for each $\eta_\nu$. It agrees well with what expected for a
coherent state with average number of photons $|\alpha|^2 \simeq
0.02$. In the inset of the Fig. 1 the frequencies $f_{\nu}$ as a
function of $\eta_{\nu}$ are compared with the theoretical
prediction $p_\nu = \exp\{-\eta_\nu |\alpha|^2\} \simeq 1-
\eta_\nu |\alpha|^2$. Notice that in this case we do not have IF
or irises in front of the detector and all the other attenuations
can be included in the generation of the state: thus the highest
quantum efficiency is taken to be $66 \%$ as declared by the
manufacturer data-sheet for the photodetector.

Finally, as a last example, we present here our preliminary data
on a single branch of PDC in pulsed regime, which is expected to
correspond to a multi-thermal state \cite{man}.

The state has been generated by pumping a 1x1x1 mm type I BBO
crystal by  a beam of a Q-switched triplicated (to 355 nm)
Neodimium-Yag laser with pulses of 5 ns, power up to 100 mJ per
pulse and 10 Hz repetition rate. Since this source was built with
the purpose of having a high spectral selection (as a step toward
a source useful for realising a quantum memory and quantum logical
gates based on Kerr effect \cite{qmk}) a monochromator with a 0.2
nm selection (obtained by an entrance 0.025 mm slit) was inserted
in the optical path.
 More in details, after having identified the precise direction of
 PDC emission at 780.2 nm by injecting into the crystal  a laser beam locked on this
 wave length at the angle such to observe stimulated emission, the
 monochromator was alligned on this beam (after a lens) and
 followed by an objective coupling the transmitted PDC to a fiber
 addressed to photo-detector.
The photo-detector was then gated by a signal coming from the
laser in order to be opened in coincidence with the pulse arrival.
On/off measurements corresponding to this window were then used
for the statistics reconstruction, whose results are shown in the
figure.

Also in this case  uncertainties in the evaluation of diagonal
element of density matrix were evaluated according to the
discussion of paragraph 3. As shown in Fig. 2 the reconstructed
statistics of our preliminary data agrees (within uncertainties)
with the one expected for a multi-thermal case. Nevertheless the
fit seems to favour a relatively small number of modes (smaller
than 10), whilst the number of modes  dictated by the ratio of the
pulse duration (5 ns) and the coherence time for the 0.2 nm
spectral selection is rather large. If one simply fixes the number
of modes to 500 or more, the fit  becomes worse (the reduced $\chi
^2$ increases from 0.5 to about 10).

The solution of this problem is postponed to a further data
acquisition and to a more careful theoretical analysis. One
possibility is that the method is reaching its limit when the
average number of photons is small (smaller than 1 in our case)
while the number of modes is very large. An alternative solution
could derive from the presence of a non PDC background, e.g. the
introduction into the fit of a strong poissonian background can
lead to a good fit (reduced $\chi ^2 \approx 3 \cdot 10^{-3}$).

For the sake of completeness, we have also considered the case of
multithermal distribution with a very low level of photons as
well. In this case the optical state  was a single branch of PDC
emission obtained by pumping a type II BBO crystal with a cw Argon
ion laser beam (351 nm) of 0.3 W. The emission was selected by a 4
nm FWHM interference filter (which fixes the coherence time of the
field) and addressed to an APD single photon detector where on/off
counts were collected in a 20 ns window. The reconstructed
statistics distribution agrees well, within uncertainties, with a
multithermal state with a large number of modes (see Fig. 2), but
at this low intensity level (0.064 photons on average) the
distinction among different optical states as coherent,
multithermal, thermal or poissonian becomes rather weak and no
clear distinction emerges from data.

\begin{figure}[h]
\begin{tabular}{ccc}

\includegraphics[width=0.3\textwidth]{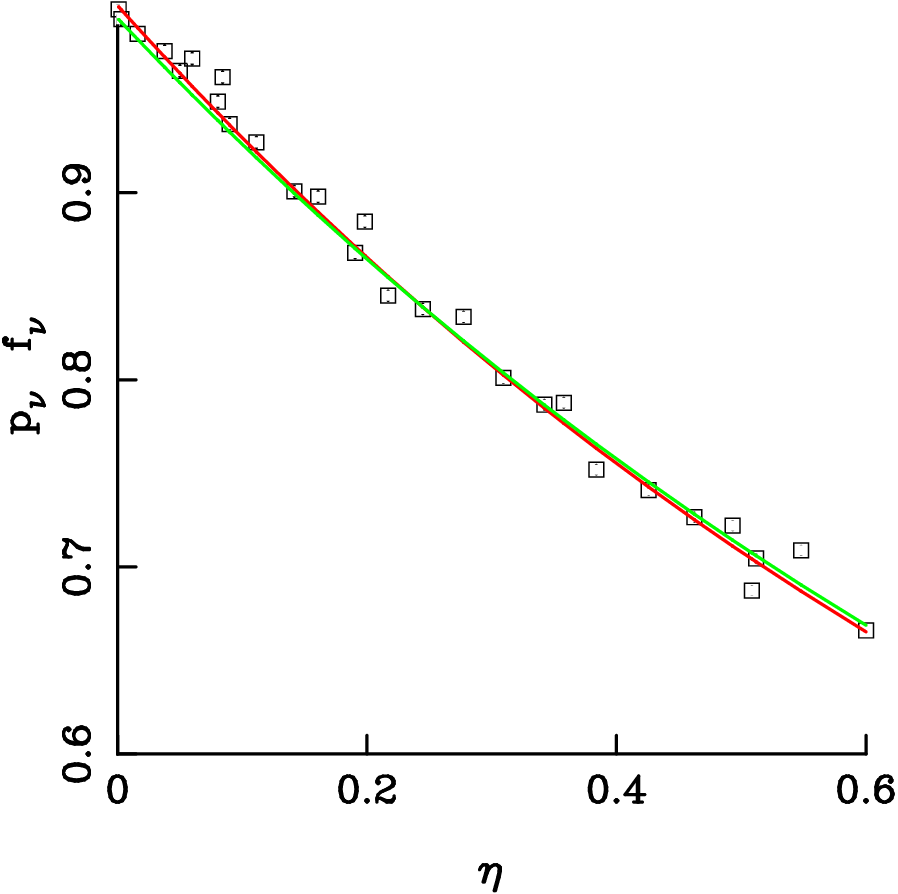}
&
\includegraphics[width=0.3\textwidth]{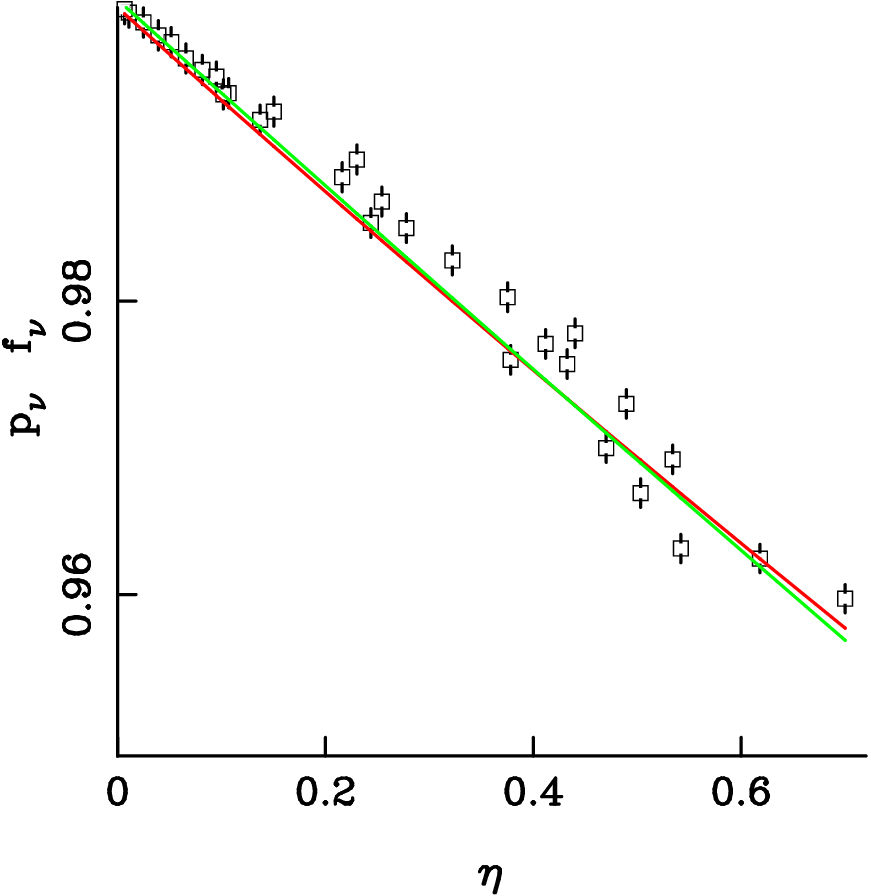}
\\

\includegraphics[width=0.3\textwidth]{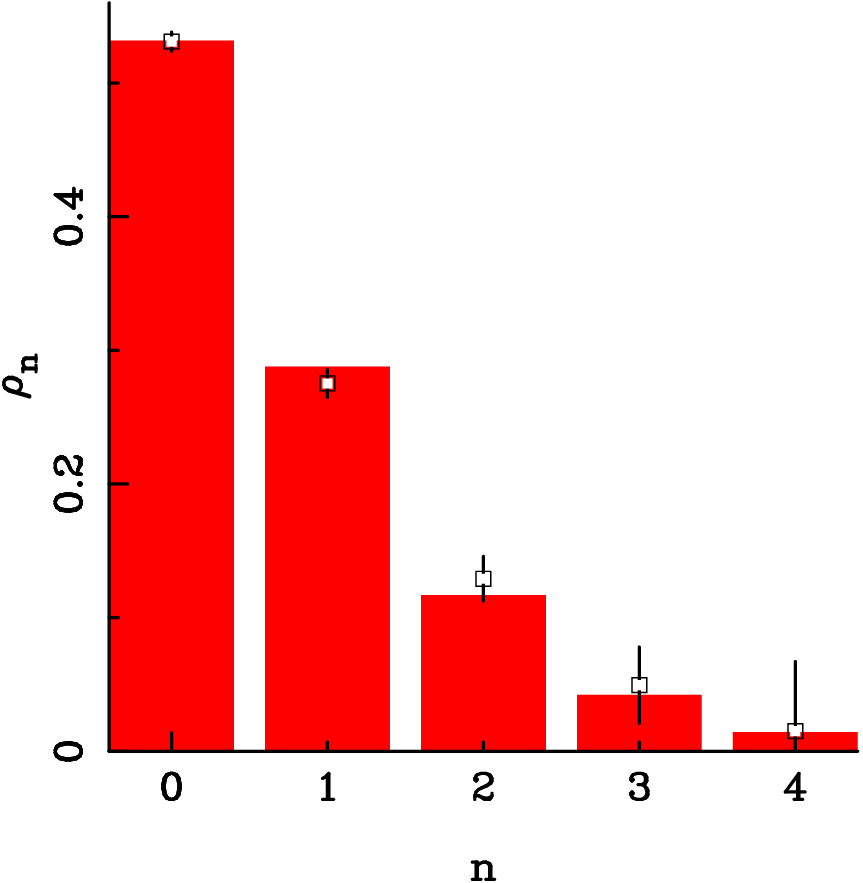}
&
\includegraphics[width=0.3\textwidth]{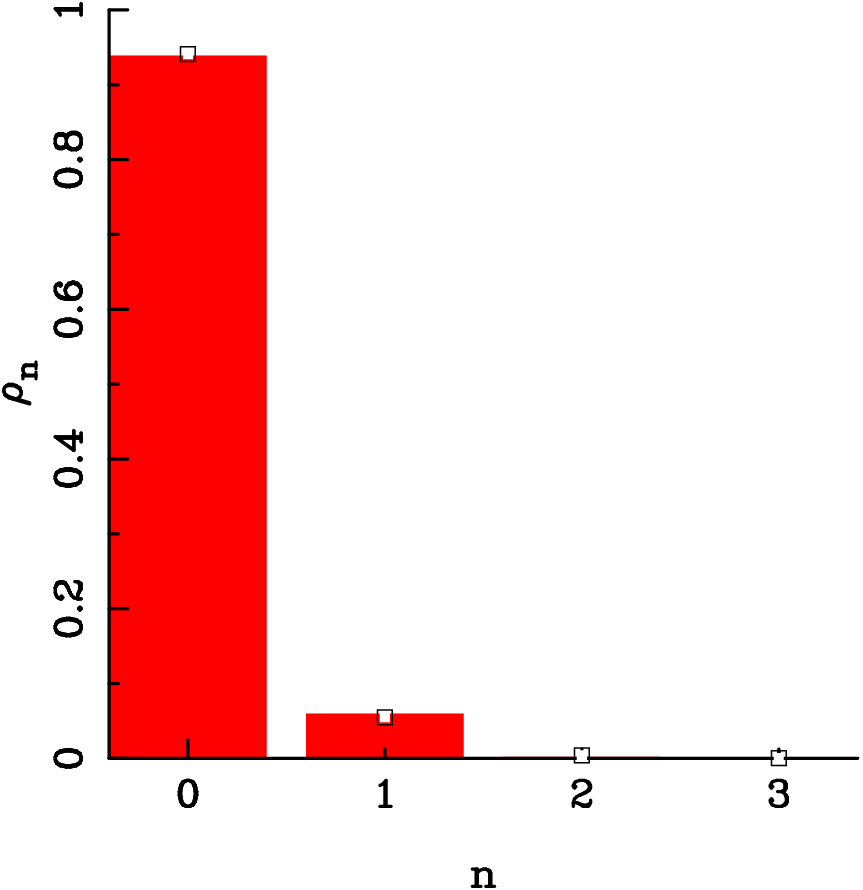}
\end{tabular}
\caption{Reconstruction of statistics of quantum optical states.
In the highest part: Experimental frequencies $f_{\nu}$ of
no-click events compared with reconstructed ones (red curve) and
theoretical model for the state (green curve). In the lowest part,
reconstructed probabilities compared with data. a) multithermal,
single branch of PDC in nanosecond pulsed regime with 0.2 nm
spectral selection. The curves correspond to 2 modes, 0.74 average
photons. b) multithermal, single branch of PDC in cw regime with 4
nm spectral selection. The curves correspond to 10000 modes 0.064
average photons. \label{f:fig2}}
\end{figure}

\section{Conclusions}

In summary in this paper we have presented  a detailed analysis of
the possibility of reconstructing experimentally diagonal elements
of the density matrix, with an accurate evaluation of
uncertainties, of quantum optical states by using on/off
detectors.

The method that has been used \cite{ar,pcount} is based on the
measurement of on/off detection frequencies for a certain optical
field when varying the quantum efficiency of the system, i.e. in
practice by interposing calibrated neutral filters on the optical
path.

Various examples have been considered extending and deepening the
analysis presented in Ref. \cite{nos}. In particular, we have
considered  PDC heralded photons, multi-thermal and attenuated
coherent states. Estimation of uncertainties with this technique
has been discussed in detail. The results of reconstruction look
very good for weak coherent and heralded photon state, while a
further analysis is still necessary for clarifying the case of
multithermal state, despite a qualitative agreement.

Altogether our results further demonstrate the potentialities of
this technique, whose main advantage with respect to alternative
schemes resides in the extreme simplicity that  allows an
extensive application for testing optical fields in a wide number
of applications.

\section{Acknowledgements}
This work has been supported by MIUR (FIRB RBAU01L5AZ-002 and
RBAU014CLC-002, PRIN 2005023443-002 and 2005024254-002), by
Regione Piemonte (E14), and by "San Paolo foundation". Turin group
thanks F. S. Cataliotti for help in locking diode laser.

\end{document}